\documentclass[useAMS,usenatbib]{mnras} 
\usepackage{graphicx}
\usepackage{amssymb, amsmath}
\usepackage{times}
\usepackage{color}
\usepackage{lscape}
\usepackage[section]{placeins}
\usepackage{breqn}

\usepackage{lineno}

\usepackage{arydshln}

\usepackage{url}

\definecolor{carrotorange}{rgb}{0.93, 0.57, 0.13}

\newcommand{\gtsima}{$\; \buildrel > \over \sim \;$}
\newcommand{\ltsima}{$\; \buildrel < \over \sim \;$}
\newcommand{\simgt}{\lower.7ex\hbox{\gtsima}}
\newcommand{\simlt}{\lower.7ex\hbox{\ltsima}}

\begin{document}


\title[Combining weak lensing and LSS]{Testing gravity on large scales by combining weak lensing with galaxy clustering using CFHTLenS and BOSS CMASS}

\author[Alam et al.] {
    Shadab Alam$^{1,2,3}$ \thanks{email: salam@roe.ac.uk}, Hironao Miyatake$^{4,5,6}$, Surhud More$^{5}$, Shirley Ho$^{1,2}$, 
\newauthor
Rachel Mandelbaum$^{1,2}$   \\
    $^{1}$ Department of Physics, Carnegie Mellon University, 5000 Forbes Ave., Pittsburgh, PA 15213 \\
    $^{2}$ McWilliams Center for Cosmology, Carnegie Mellon University, 5000 Forbes Ave., Pittsburgh, PA 15213 \\
    $^{3}$ Institute for Astronomy, University of Edinburgh, Royal Observatory, Blackford Hill, Edinburgh, EH9 3HJ , UK \\
    $^{4}$Jet Propulsion Laboratory, California Institute of Technology, Pasadena, CA 91109 \\
    $^{5}$ Kavli Institute for the Physics and Mathematics of the
    Universe (WPI), UTIAS, The University of Tokyo, Chiba, 277-8583,
    Japan\\
    $^{6}$ Department of Astrophysical Sciences, Princeton University, Peyton Hall, Princeton NJ 08544, USA 
}

\date{\today}
\pagerange{\pageref{firstpage}--\pageref{lastpage}}   \pubyear{2015}
\maketitle
\label{firstpage}

\begin{abstract}
We measure a combination of gravitational lensing, galaxy clustering, and redshift-space distortions called $E_G$. The quantity
$E_G$  probes both parts of metric potential and is insensitive to galaxy bias and $\sigma_8$. These
properties make it an attractive statistic to test $\Lambda$CDM, General Relativity and its
alternate theories.  We have combined CMASS DR11 with CFHTLenS and recent measurements of $\beta$
from RSD analysis, and find $E_G(z = 0.57) = 0.42 \pm 0.056$, an 13\% measurement in agreement with
the prediction of general relativity $E_G(z = 0.57) = 0.396 \pm 0.011$ using the Planck 2015
cosmological parameters. We have corrected our measurement for various observational and theoretical
systematics. 
Our measurement is consistent with the first measurement of $E_G$ using CMB lensing in place of
galaxy lensing \citep{Pullen2015data} at small scales, but shows 2.8$\sigma$ tension when compared with their final results including large scales. This analysis with future surveys will provide improved statistical error and better control over systematics to test General Relativity and its alternate theories.
\end{abstract}



\section{Introduction}
\label{sec:intro}

The theory of General Relativity (GR) is the most successful theory of the gravity. The GR was first proposed by \citet{Einstein1915}. 
GR has passed the most stringent tests at solar system scales \citep{Sakstein2015}. But it is still an ongoing pursuit to test the predictions of GR at cosmological scale before we finally declare that it is the ultimate theory of gravity. There are some  observational mysteries like dark matter \citep{Zwicky1937, Kahn1959, Rubin1970} and dark energy \citep{Riess1998,Perlmutter1999} which cannot be explained with the current models. But if one ignores the questions about origin of dark matter and dark energy, then $\Lambda$CDM-GR is in good agreement  with Cosmic Microwave Background (CMB) \citep{Wmap2013, PlanckI}, Baryon Acoustic Oscillation (BAO) \citep{Eis2005,Cole2005,Hutsi2006,Kazin2010, Percival2010, And13, Anderson2014} and Hubble constant \citep{Riess2011}. One of the fundamental theoretical mysteries is the incompatible nature of quantum mechanics and GR. The nature of time in the two theories is so different that it is difficult to combine them in a single framework \citep{Unruh1993,Anderson2010}.  In order to further the understanding of these mysteries and develop consistent theories, it is important to test the predictions of these theories in various regimes. A fundamental difficulty of testing modifications to GR is the ability to absorb these modifications in dark energy. Fortunately, modified gravity predicts large scale structures different from those predicted by Einstein's theory of gravity \citep{Koyama2006}.

GR predicts many signatures of structure formation which can be observed in a wide variety of surveys. Two complimentary signals measured are weak gravitational lensing and redshift space
distortions (RSD). Gravitational lensing was first proposed by Einstein in \citet{Einstein1916}. Weak gravitational lensing is a statistical measurement of deflection of photons due to gravitational interaction with the matter density \citep[for a review, see][]{Bartelmann2001, Refregier2003, Schneider2005, Hoekstra2008, Massey2010, Weinberg2013}. Its signal is imprinted in the cross-correlation of background galaxy shapes with foreground galaxy positions
\citep[e.g.,][]{2012ApJ...744..159L,2013MNRAS.432.1544M,2014MNRAS.437.2111V,2015MNRAS.446.1356H,2015MNRAS.447..298H,2015arXiv150502781Z},
and can be measured as ``cosmic shear'' (the auto- and cross-correlation of pairs of galaxy shapes; e.g., \citealt{2013MNRAS.432.2433H,2013ApJ...765...74J}).
The redshift space distortion is the measurement of anisotropy produced in the galaxy auto-correlation
function due to the peculiar velocity component in the galaxy redshift.  This anisotropy allows us
to measure the growth rate ($f=d\ln D/d\ln a$) of cosmic structure formation. It was first
introduced by \cite{Kaiser87} and then further developed by others \citep{Hamilton92,Socco2004}. It
has been measured by various galaxy redshift surveys using different modeling schemes
\citep{2dFGRS, WiggleZ, 6dFGRS, Vipers, Chuang13, Ariel13, Beutler13, Alam2015}.

The larger surveys and novel combinations of probes will test the predictions of GR with unprecedented precision.  One such combination of redshift space distortion and weak gravitational
lensing was proposed by \cite{Zhang2007}. It is important to test the relative amplitude of the
effect of RSD and weak gravitational lensing as it probes space and time both parts of the
metric. They have constructed a quantity $E_G$ which can be measured by combining the signal from
these two complimentary measurements. It has been proposed that  $E_G$ has the potential to serve as
the most precise signal to test the nature of gravity \citep{Zhang2007}. $E_G$ is independent of
linear bias and the amplitude of matter fluctuations ($\sigma_8$).  \cite{Reyes2010} has measured
the first signal of $E_G$ using a lower redshift sample from the Sloan Digital Sky Survey at an effective redshift of 0.32. Recently \cite{Blake2015} reported the measurement of $E_G$ at two different redshifts, 0.32 and 0.57. A number of possible theoretical systematics of $E_G$ is discussed in \cite{Leonard2015}.

In this paper we measure $E_G$ by combining the measurement of the weak gravitational lensing
\citep{Miyatake:2015} from the Canada-France-Hawaii Lensing Survey \citep{Heymans2012}, hereafter
referred to as CFHTLenS, with the measurement of redshift-space galaxy clustering  from the Data Release
11 (DR11) CMASS sample \citep{Alam2014} of  Baryon Oscillation Spectroscopic Survey (BOSS;
\citealt{Ahn2012}), which is part of Sloan Digital Sky Survey III (SDSS-III; \citealt{Eisenstein2011}). 

We have organized this paper in the following manner. In section \ref{sec:theory}, we provide some brief theoretical background of the $E_G$. Section \ref{sec:data} describes the samples of data used in our measurements. Section \ref{sec:measurement} describes the measurement of different components of $E_G$ in detail with some systematic corrections. Section \ref{sec:nbody} provides the details of $N$-body simulation used in our analysis. The list of possible systematics affecting our $E_G$
measurement with possible corrections is discussed in section \ref{sec:systematic}. Finaly, we
provide our main measurement and estimate of uncertainity on the measurement in section
\ref{sec:results}. We end our paper with the discussion of the main points of our analysis and the implications of our results, along with some future directions, in section \ref{sec:discussion}. Our fiducial cosmology is flat $\Lambda$CDM with $\Omega_m=0.31 $ and $h=0.67$ all throughout the paper unless mentioned otherwise.

\section{Theory}
\label{sec:theory}

The combination of galaxy-galaxy clustering, redshift space distortions and galaxy-galaxy lensing
provides $E_G$. The measurements of lensing and clustering signals have been transformed to new
quantities called $\Upsilon$ in order to reduce the impact of theoretical uncertainties and failures
of certain approximations on small scales (as discussed later). The combined probe $E_G$ has been operationally  defined in
\citet{Reyes2010} as follows:

\begin{equation}
\label{eq:EGobservation}
 E_G(r_p)= \frac{\Upsilon_{gm}(r_p)}{\beta \Upsilon_{gg}(r_p)} 
 \end{equation}
where $\beta=f/b$ is the redshift space distortion parameter with $f$ being logarithmic derivative
of growth with respect to scale factor and $b$ is the linear bias. The quantities $\Upsilon_{gm}$
and $\Upsilon_{gg}$ are called  galaxy-matter and galaxy-galaxy annular differential surface
densities respectively \citep[ADSDs;][]{Baldauf2010}.  $\Upsilon_{gm}$ is defined as
\begin{align}
\label{eq:Ups-gm}
\Upsilon_{gm} & ( r_p) =   \Delta \Sigma_{gm}(r_p) - \left(\frac{R_0}{r_p}\right)^2  \Delta \Sigma_{gm}(R_0) \nonumber\\
   = & \frac{2}{r_p^2} \int_{R_0}^{r_p} dR^\prime R^\prime \Sigma_{gm}(R^\prime)-\Sigma_{gm}(r_p) + \left(\frac{R_0}{r_p}\right)^2 \Sigma_{gm}(R_0). \nonumber \\
 \end{align}
The observable for the weak gravitational lensing is the sum of the tangential shear from lensing
($\gamma^G_t$) and galaxy intrinsic shear ($\gamma^I$). Assuming galaxy intrinsic shear is
negligible, lensing observation is proportional to $\Delta \Sigma(r_p)=\bar{\Sigma}(<r_p) -
\Sigma(r_p)$, which is a measure of excess surface mass density. The value of $\Delta \Sigma(r_p)$
depends on all scales below $r_p$, which is not quite well described by linear
theory. $\Upsilon_{gm}$, shown in Eq.~\eqref{eq:Ups-gm}, is an attempt to cast the lensing observable
$\Delta \Sigma(r_p)$ in such a way that it becomes independent of information below a certain scale
$R_0$.  $\Upsilon_{gg}$ is defined as
\begin{multline}
\label{eq:Ups-gg}
 \Upsilon_{gg}(r_p)  =  \rho_{crit} \biggl[\frac{2}{{r_p}^2} \int_{R_0}^{r_p} dR^\prime R^\prime w_{gg}(R^\prime) - \\ 
 w_{gg}(r_p) +  \left(\frac{R_0}{r_p}\right)^2 w_{gg}(R_0) \biggr] 
 \end{multline}
Here $w_{gg}$ represents the projected galaxy-galaxy correlation function. 
These definitions ensure that despite measuring slightly different observables for the lensing and
clustering, they are transformed to the same statistic of the correlation function, so that the
theoretical prediction of $E_G$ is equivalent to the measurement.  
Theoretically $E_G$ can be defined in terms of metric perturbations:
 \begin{equation}
 E_G = \frac{\nabla^2 (\Psi(r)-\Phi(r))}{3H_o^2 a^{-1} \theta} \,,
 \label{eq:EGtheory}
 \end{equation} 
where $\theta$ is the perturbation in matter velocity field, $H_0$ is the Hubble parameter today and
$a$ is the scale factor. The $\psi$ and $\phi$ represent metric perturbations to the time and space
components, respectively, assuming a Friedmann-Robertson-Walker (FRW) metric with a flat universe.  The numerator $\nabla^2 (\Psi(r)-\Phi(r))$ probes the lensing convergence and the $\theta$ in denominator probes the redshift space distortions. As shown in \cite{Hojjati2011}, the time-time and momentum Einstein field equation in GR, under the assumption of negligible anisotropic stress and non-relativistic matter species, becomes the simple algebraic equation,
\begin{align}
k^2\Psi =& -4\pi G a^2 \rho(a) \delta \,, \\ 
\Phi=&-\Psi \, ,
\end{align}
where $\rho$ is the background matter density and $\delta$ is the matter density perturbation. In modified theories of gravity these relations are different, which requires two extra functions $\mu(k,a)$ and $\gamma(k,a)$ to account for departure from GR \citep{Hojjati2011}.
\begin{align}
k^2\Phi =&-4\pi G a^2 \mu(k,a) \rho \Delta \\
\Phi =& - \gamma(k,a) \Psi \, ,
\label{eqn:modGR}
\end{align}

The perturbation equations in $\Psi$ and $\Phi$ are in fourier space which should be related to its real space counter part $\Psi^{\prime}(r) =\Psi(k) e^{-ikr},\, \Phi^{\prime}(r) =\Phi(k) e^{-ikr}$ . This gives us the relation $\nabla^2 (\Psi^{\prime}(r)-\Phi^{\prime}(r)) =-k^2 (\Psi(k)-\Phi(k))$. We combine the perturbation equation and the definition of $E_G$ with  $\Omega_M(z=0)= \frac{8\pi G \rho_o}{3 H_o^2}$,  $\rho =\rho_o a^{-3}$ and $\theta=-f\delta$ to get our theoretical prediction of $E_G =-\Omega_M(z=0) \mu(k,a) (\gamma(k,a)+1)/2f$. The $E_G= \Omega_M(z=0)/f$ can be recovered for GR by substituting $\mu=-1$ and $\gamma=1$.

It is non-trivial to see the connection between our theoretical definition (Eq.~\ref{eq:EGtheory}) and the observational definition (Eq.~\ref{eq:EGobservation}). We provide a brief outline to make this connection a little bit easier. Please refer to \citet{Reyes2010} and \citet{Baldauf2010} for more details. The statistics $\Upsilon$ for galaxy-matter and galaxy-galaxy can be written in terms of their corresponding power spectrum as follows.
\begin{equation}
\Upsilon_{gg,gm}(r_p;R_0)= \rho_{\rm crit} \int P_{gg,gm} W_{\Upsilon}(k;r_p,R_0) dk \,,
\end{equation}
where $W_{\Upsilon}(k;r_p,R_0)$ is the window function for $\Upsilon$ given in equation 17 of \citet{Baldauf2010}. We know that galaxy-matter power spectrum is proportional to the cross power of convergence ($\nabla^2 (\Psi-\Phi)$) and galaxy. This implies that $P_{gm} \equiv \left\langle\nabla^2 (\Psi-\Phi) \delta_g\right\rangle$. We also know that galaxy-galaxy power spectrum can be written as follows,
\begin{equation}
P_{gg} \equiv \left\langle \delta_g \delta_g \right\rangle 
= -\frac{1}{\beta}\left\langle \theta \delta_g \right\rangle
\end{equation}
The first equivalence is the definition of the galaxy-galaxy power spectrum. The second equality results from the fact that for linear regime, matter conservation relates velocity perturbations ($\theta$) to matter perturbations ($\delta$) by $\theta=-f\delta$  and, linear bias model relates $\delta$ to $\delta_g$ by $\delta_g=b \delta$. Now, the ratio of $\Upsilon_{gm}$ and  $\beta\Upsilon_{gg}$ gives the ratio of power spectrum, which will be proportional to $(\nabla^2 (\Psi-\Phi) \delta_g)/\theta \delta_g$ . Therefore our observational definition given by Eq.~\eqref{eq:EGobservation} is same as the theoretical definition given by Eq.~\eqref{eq:EGtheory}.

\section{DATA} 
\label{sec:data} 

We use the SDSS-III BOSS CMASS sample and shape measurements from CFHTLenS data to measure the galaxy-galaxy clustering, galaxy matter cross-correlation and redshift space distortions parameter. We describe the data sets used in our analysis in the following sections.

\subsection{BOSS CMASS}
\label{sec:boss} 

We use data included in data release 11 (DR11; \citealt{Alam2014}) of the Sloan Digital Sky Survey (SDSS; \citealt{York2000}). Together, SDSS I, II \citep{Abazajian2009} and III \citep{Eisenstein2011} used a drift-scanning mosaic CCD camera \citep{Gunn1998} to image over one-third of the sky (14555 square degrees) in five photometric bandpasses \citep{Fukugita1996,Smith2002,Doi2010} to a limiting magnitude of $r <22.5$ using the dedicated 2.5-m Sloan Telescope \citep{Gunn2006} located at the Apache Point Observatory in New Mexico. The imaging data were processed through a series of pipelines that perform astrometric calibration \citep{Pier2003}, photometric reduction \citep{Lupton1999}, and photometric calibration \citep{Padmanabhan2008}. All of the imaging was reprocessed as part of SDSS Data Release 8 (DR8; \citealt{Aihara2011}).
BOSS \citep{Dawson2013} is designed to obtain spectra and redshifts for 1.35 million galaxies over a footprint covering 10,000 square degrees. These galaxies are selected from the SDSS DR8 imaging and are being observed together with 160,000 quasars and approximately 100,000 ancillary targets. The targets are assigned to tiles using a tiling algorithm that is adaptive to the density of targets on the sky \citep{Blanton2003}. Spectra are obtained using the double-armed BOSS spectrographs \citep{Smee2013}. Each observation is performed in a series of 900-second exposures, integrating until a minimum signal-to-noise ratio is achieved for the faint galaxy targets. This ensures a homogeneous data set with a high redshift completeness of more than 97\% over the full survey footprint. Redshifts are extracted from the spectra using the methods described in \citet{Bolton2012}. A summary of the survey design appears in \citet{Eisenstein2011}, and a full description is provided in \citet{Dawson2013}.

We use the CMASS sample of galaxies  \citep{Bolton2012} from Data Release 11. The CMASS sample has 690,826  massive galaxies covering 8498 square degrees in the redshift range $0.43<z<0.70$, which correspond to an effective volume of 6 Gpc$^{3}$.

\subsection{CFHTLENS}

For the galaxy-galaxy lensing measurements, we use
the deeper and better quality imaging data from the Canada France
Hawaii Telescope Legacy survey (CFHTLS). This data allows us to
measure the tangential distortion of background galaxies around our
sample of CMASS galaxies. We use the photometric reduction and image
shape determinations in the publicly available CFHTLenS
catalog\footnote{\url{http://www.cfhtlens.org/astronomers/data-store}}.
The quantities needed for each galaxy, namely its shear estimate,
shear calibration factors, weight, and the posterior distribution of
its photometric redshift distribution are provided in the catalog
\citep{Heymans2012, Erben2013, Miller:2013, Hildebrandt:2012}. We use
the same quality cuts on the data as were applied in
\citet{Miyatake:2015}. Finally we note that the overlap between the
CFHTLS and the DR11 BOSS fields is limited to an area of about 105
deg$^2$. The number of CMASS galaxies that lie within the CFHTLS
footprint is $8899$.

\section{Measurements}
\label{sec:measurement}
The quantity $E_G$ is a combination of galaxy-galaxy annular differential surface density
($\Upsilon_{gg}$), galaxy-matter annular differential surface density ($\Upsilon_{gm}$) and the redshift space distortion parameter ($\beta$). In the following section we describe the procedure to obtain each of these signals.

\subsection{Galaxy-Galaxy annular surface density ($\Upsilon_{gg}$)}
\label{sec:ups_gg}

The data from SDSS-III BOSS includes the three-dimensional positional
information of CMASS galaxies, which enables us to perform a high
signal-to-noise measurement of the projected correlation function, $w_{\rm
p}$, and the associated Galaxy-Galaxy annular surface density,
$\Upsilon_{gg}$. We account for a number of subtle selection effects in order to
obtain a precise measurement of clustering \citep{Ross:2012}. The spectroscopic
target sample is obtained from the SDSS imaging observations after the
application of a variety of colour and photometric selection cuts
\citep{Dawson2013,Reid2016}.  However, due to the limited number of fibers available,
not all galaxies from this target sample can be allocated a fiber while
performing spectroscopic observations to determine their redshifts. This could
also happen if two targets are within $62''$ of each other and hence cannot
be simultaneously observed due to the finite size of fibers. If such
fiber-collided galaxies lie in a region of the sky which is visited multiple
times (due to overlaps in the target tiling) then they may have redshift
measurements. There are also instances where a galaxy is assigned to a fiber,
but its redshift could not be obtained. Finally, there are also instances where
it is difficult to perform star-galaxy separation, especially in fields with a
high number density of stars.  These effects have been quantified in the parent
DR11 catalog of CMASS galaxies by assigning a weight to each galaxy such that
\begin{equation}
w_{\rm l} = w_\ast w_{\rm see}\, ( w_{\rm noz} + w_{\rm cp} - 1 ) \,,
\end{equation}
where $w_{\rm noz}$ is the weight assigned to a galaxy if it is the nearest
neighbour (in the plane of the sky) of a redshift failure galaxy, $w_{\rm cp}$
is similarly assigned to account for the nearest neighbours of fiber collided
galaxies\footnote{Nearest neighbour corrections have been shown to accurately
correct for fiber collisions above the fiber collision scale ($\sim 0.4 h^{-1}$ Mpc)
by \citet{Guo:2012}. Both $w_{\rm noz}$ and $w_{\rm cp}$ are equal to unity by
default for all galaxies. Their values are incremented for the nearest
neighbours of every redshift failure or fiber collided galaxy.}, and $w_\ast$ and $w_{\rm see}$
accounts for the systematic relationship between the density of stars and seeing (respectively) with
the density of
BOSS target galaxies \citep[for details, see][]{Anderson2014}. The BOSS parent
catalog contains an additional weight, $w_{\rm FKP}$, for each galaxy which
depends upon the number density of galaxies in the sample at its redshift
\citep{Feldman:1994}. The total weight for each galaxy that we use is given by
\begin{equation}
w_{\rm tot}= w_{\rm l} w_{\rm fkp}\,.
\label{eq:cmass_weight}
\end{equation}
We use catalogues of random points with the same angular and redshift selection
as our galaxy subsample. These random catalogs consist of about 50 times more
points than the number of galaxies in each of our subsamples. We assign each
random point a weight of $N_{\rm gal}/N_{\rm ran}$ to account for this
difference. In practice, we use the random catalogs provided with SDSS DR11
\citep{Reid2016}.

We measure the correlation function of galaxies, $\xi(r_p, \Pi)$, where
$r_p$ is the projected separation of galaxies, and $\Pi$, their
line-of-sight separation, using the estimator proposed by
\citet{LandySzalay93},
\begin{equation}
\xi(r_p,\Pi)=\frac{DD-2DR+RR}{RR}\,.
\end{equation}
Here, $DD$, $RR$ and $DR$ represent the number of appropriately weighted pairs
of galaxies with a given separation $(r_p,\Pi)$, where both galaxies lie
either in the galaxy catalog or the random catalog or one in each of the
catalogs, respectively. The projected correlation function is obtained by
integrating $\xi(r_p, \Pi)$ along the line of sight,
\begin{equation}
w_{\rm p}(r_p)=2 \int_{0}^{\Pi_{\rm max}} \xi(r_p,\Pi) \,d \Pi\,,
\label{eq:wp}
\end{equation}
where we adopt $\Pi_{\rm max}=100 h^{-1}$Mpc. We then convert the projected correlation function into galaxy-galaxy annular differential surface density following Eq.~\eqref{eq:EGobservation}, where we adopt $R_0=1.49 h^{-1}$Mpc. Figure \ref{fig:wp-ups-sim} shows the projected correlation function and  galaxy-galaxy ADSDs measured from data and simulation. We carry out this measurement at $1.2 < r_p < 47~h^{-1}{\rm Mpc}$ divided into
9 bins.

\subsection{Galaxy-Matter annular surface density ($\Upsilon_{gm}$)}
\label{sec:upgm}
For the weak lensing measurement, we followed the procedure described in
\cite{Miyatake:2015}. In this paper we summarize the
procedure; we encourage those who are interested in details to read reference.
The tangential shear caused by lensing is related to the excess
surface mass density as
\begin{equation}
 \gamma^G_t=\frac{\Delta\Sigma(r_p)}{\Sigma_{\rm cr}},
\end{equation}
where $\Sigma_{\rm cr}$ is defined as
\begin{equation}
 \Sigma_{\rm cr}(z_l, z_s) = \frac{c^2}{4\pi G}\frac{d_A(z_s)}{d_A(z_l)d_A(z_l, z_s)(1+z_l)^2}.
 \label{eq:sig_cr}
\end{equation}
Here, $d_A(z_l)$, $d_A(z_s)$, and $d_A(z_l, z_s)$ are the angular
diameter distance to lens, source and between lens and source. The
factor of $(1+z_l)^{-2}$ is due to our use of comoving coordinates. Using
lens-source pairs, the excess surface mass density is calculated as
\begin{equation}
\Delta\Sigma(r_p) = \frac{\sum_{ls} w_{ls} e_t^{ls} \Sigma_{\rm
 cr}^{ls}}{(1+K(r_p))\sum_{ls} w_{ls}},
\end{equation}
where $e_t$ is the ellipticity of a source galaxy given by the
CFHTLenS catalog\footnote{The ellipticity in the CFHTLenS catalog is
  defined by $|e|=(a-b)/(a+b)$, where $a$ and $b$ are the semi-major and semi-minor axes
  of the ellipse.  The ensemble expectation value of this ellipticity definition is equal to the
  lensing shear.}. When calculating $\Sigma_{\rm
  cr}^{ls}$, we use the probability distribution function of
photometric redshift (photo-$z$). We use the weight $w_{ls}=w_{{\rm tot},l} w_s
\Sigma_{\rm cr}^{-2}$, where $w_{{\rm tot},l}$ is the weight of each
lens galaxy given by Eq.~\eqref{eq:cmass_weight} and $w_s$ is the
weight of each galaxy given by the CFHTLenS catalog. The factor of
$(1+K(r_p))^{-1}$ is calculated using the multiplicative bias correction
factor given by the CFHTLenS catalog. We use the same $r_p$ binning as the clustering measurement ($\Upsilon_{gg}$).

We perform two systematic tests for the lensing measurement. The first
is a test for contamination from galaxies that are physically associated with
lens galaxies, and therefore not lensed. If we wrongly select those galaxies as sources,
the lensing signal is diluted. This effect can be diagnosed using the
so-called ``boost factor'' which is a ratio of the sum of the weight
of galaxies behind lens galaxies to that behind random points \citep{Sheldon2004}. If the lensing signal is diluted, the boost factor is larger than unity. In our measurement, we find the boost factor is consistent with unity within $1.6\%$ at $r_p \simgt 1.5~h^{-1}{\rm Mpc}$, the scales used for this study. The statistical error in the boost factor is subdominant compared to the statistical error coming from the shape noise.  
The second systematic test is for the effect of imperfect PSF correction. This can be diagnosed by
measuring the lensing signal
around random points, which exhibits a spurious signal for certain types of imperfection in the PSF
correction. In our measurement, the lensing signal around random points deviates from zero at $r_p
\simgt 5~h^{-1}{\rm Mpc}$. We find that the 45-degree rotated signal, which
should be consistent with zero around galaxies, deviates from zero at these scales. After
subtracting the signal around random points, the 45-degree rotated signal becomes consistent with zero except for the outermost bin at $r_p\sim40~h^{-1}{\rm
  Mpc}$. Thus we apply the same correction to the lensing signal, and
discard the outermost bin. The correction ranges from 5\% to 14\% of the lensing signal before correction. The statistical uncertainty on this correction is very small since the number of random points is much larger than the number of real lenses.  This correction is a valid way to correct for shear systematics that were not fully removed by the PSF correction routine, provided that the source of the systematics does not correlate with the lens number density.  Since the lenses are selected in one survey and the shears measured in another, there is no reason for such a correlation to exist, so the correction is valid and we do not associate a systematic uncertainty with this correction \citep{2005MNRAS.361.1287M}. 

\begin{figure}
\includegraphics[width=0.45\textwidth]{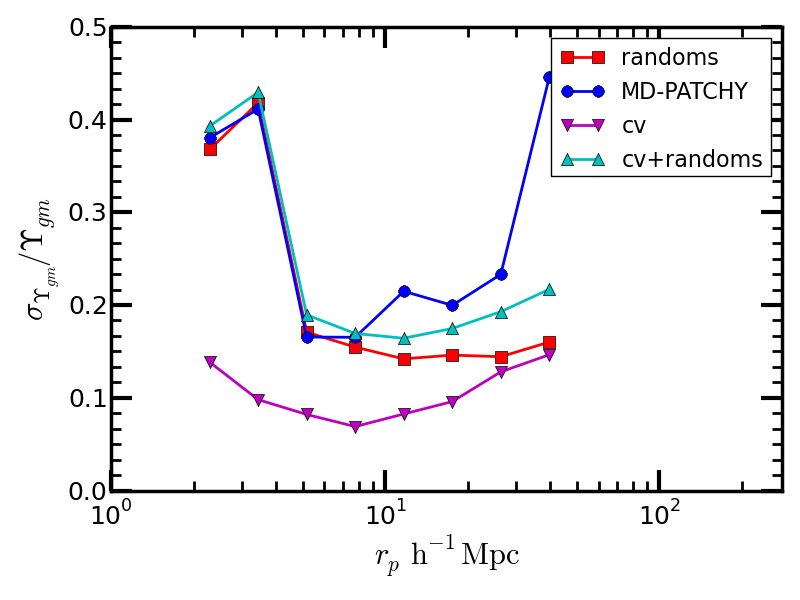}
\caption{The plot shows the fractional error in galaxy-matter annular surface density
  ($\Upsilon_{gm}$) as a function of $r_p$. The blue circles show the error estimate using MD-Patchy
  mocks, the red squares show the error estimates from random lenses, the magenta triangles are
  estimates of the cosmic variance contribution from $N$-body simulations, and the cyan triangles
  are obtained by adding cosmic variance and the error from random lenses in quadrature. We use the error estimate from MD-Patchy in our final results.}
\label{fig:Lensing-error}
\end{figure}

When calculating the covariance of the lensing signal, we must account for the correlated shape
noise, lens shot noise, and cosmic variance. The correlated shape noise is caused by the use of the
same source galaxies multiple times, since the stacking annuli overlap for different lens
galaxies. The lens shot noise stems from the noise in the distribution of redshift or other
properties of CMASS galaxies in the CFHTLenS field due to the small area compared to the the full
BOSS sample. The cosmic variance is due to the large-scale fluctuation modes larger than a survey
region. To account for them, we estimate the covariance matrix using 150 realizations of the lensing
signal around MD-Patchy mocks \citep{2016MNRAS.456.4156K}. We compare this covariance to that
estimated from the lensing signal around random points, which is considered to have just part of the
correlated shape noise\footnote{Clustered lenses have more correlated shape noise than random
  points.}, and find that the lens shot noise and cosmic variance make up about the half of the
covariance at $r_p\simgt 10h^{-1}{\rm Mpc}$. We also calculate the expected covariance by adding the
random covariance to the lens shot noise and cosmic variance estimated from $N$-body simulations
(see Section~\ref{sec:cv} for details). We confirm that the difference between the covariance
estimated from MD-Patchy mocks and the expected covariance is within 3 per cent at $r_p\simlt
30h^{-1}{\rm Mpc}$. Figure~\ref{fig:Lensing-error} shows a comparison  of the lensing error
estimates from different sources. We use the error estimate from MD-Patchy in our final results.

\begin{figure}
\includegraphics[width=0.45\textwidth]{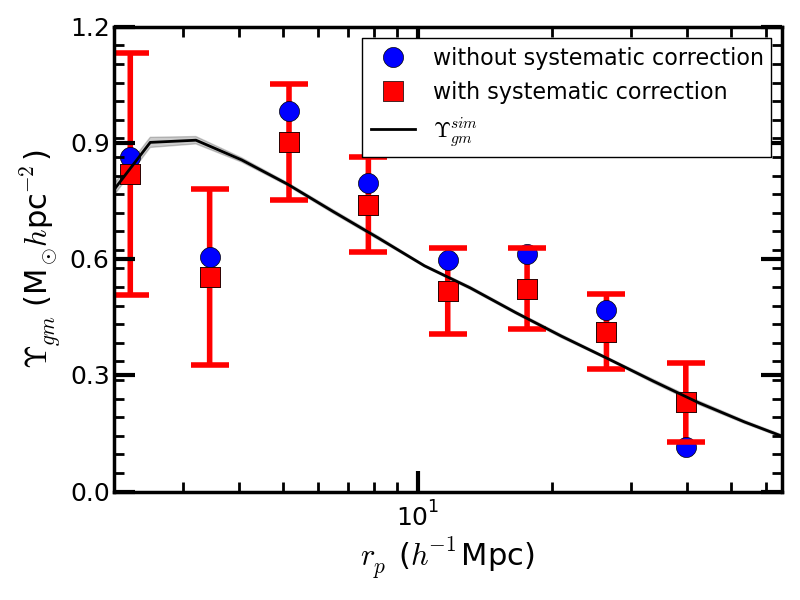}
\caption{The plot shows the galaxy-matter annular surface density ($\Upsilon_{gm}$) measured from
  CFHTLenS and CMASS catalogues. The blue points are measurements before and red points are after
  applying an additional systematic correction for imperfect PSF correction resulting in additive
  shear systematics. The black line and shaded region shows the measurement of $\Upsilon_{gm}$ and
  1$\sigma$ error from an $N$-body simulation.}
\label{fig:ups-gm-data}
\end{figure}

We then convert the excess surface mass density and its covariance into galaxy-matter
annular differential surface density following Eq.~\eqref{eq:Ups-gm}.
Figure~\ref{fig:ups-gm-data} shows our measured galaxy-matter annular differential surface density,
including the size of the systematic correction for imperfect PSF correction.

\subsection{Redshift Space Distortions parameter ($\beta$)}
We measure the two-dimensional auto-correlation function of the BOSS CMASS galaxies using the Landy-Szalay
\citep{LandySzalay93} estimator. The correlation function is first binned in ($r,\mu$), where $r$ is
the three-dimensional galaxy pair separation and $\mu=\cos(\theta)$ with $\theta$ being the angle made by the pair of
galaxies from the line of sight. The galaxy-galaxy auto-correlation is projected onto the Legendre basis in order to obtain the monopole ($\xi_0$) and quadruple ($\xi_2$) moments.  The monopole and quadruple moments of the correlation function are evaluated between 6~to 198~$h^{-1}$Mpc in linear bins of width 8~$h^{-1}$Mpc. The bin size of 8~$h^{-1}$Mpc is chosen to optimize signal-to-noise
without smoothing out the important physics. We have used 600 PTHalo mocks \citep{Manera13} to generate an estimate of the covariance matrix for the measured correlation function. The fit to the monopole and quadruple moments of the correlation function is obtained using Convolution Lagrangian
Perturbation Theory (CLPT) and Gaussian Streaming Model (GSM) \citep{Carlson12,Wang13}. 

The theoretical model has been tested using PTHalo mocks. It was shown that our model gives accurate
prediction of $\xi_{0,2}$ at scales ranging from 30 $h^{-1}$Mpc to 126 $h^{-1}$Mpc with 8 $h^{-1}$Mpc bin width.  The measured $f\sigma_8(z=0.57)=0.462\pm0.041$ and $b\sigma_8(z=0.57)=1.194\pm0.032$, as reported in
\cite{Alam2015} .
The RSD parameter $\beta$ is computed by taking the ratio of the measured growth rate
$f$ and bias $b$, i.e., $\beta = f/b$. This gives us $\beta(z=0.57)=0.387\pm0.042$, while accounting
for the correlation between growth rate and bias. The complete redshift space distortion analysis is
reported in \cite{Alam2015}, including the list of parameters marginalized and prior used on those
parameters in table 2 of that work. A comparison of this measurement with other similar measurements is shown in Figure 6 of \cite{Alam2015}.

We note that the measurement of $\beta$ is obtained using scales above 30 $h^{-1}$Mpc whereas our final $E_G$ measurement use scales below 30 $h^{-1}$Mpc. Ideally one would want to measure $\beta$ using same scales. But, unfortunately, the modeling used in the current measurement of $\beta$ is not good enough to extend to smaller scales. The $\beta$ consists of two quantities, growth rate and bias. We do account for the fact that bias will be scale dependent and different at smaller scale compared to large scale measurement through a correction factor $C_b$ (see section \ref{sec:cb} for details). But we have an inherent assumption that the growth rate measured using larger scales are constant and applicable for smaller scales. This makes the current measurements of $E_G$ slightly weaker than its full potential, which should be improved upon in the future measurements with better RSD modeling.

\section{$N$-body Simulations} 
\label{sec:nbody}

\begin{figure}
\includegraphics[width=0.5\textwidth]{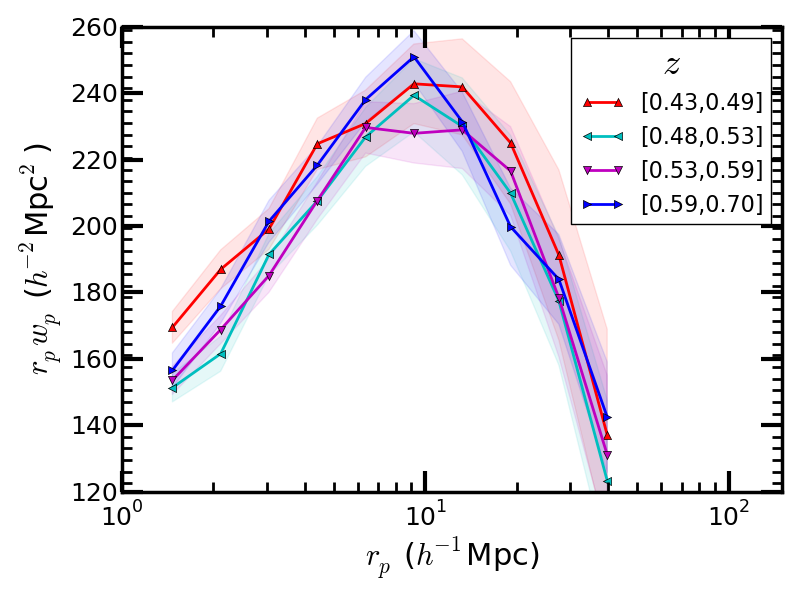}
\caption{The CMASS sample is divided into four redshift bins, each containing roughly equal number
  of galaxies. We show the $w_p$ for each of the redshift bins, with the shaded region showing the
  $1\sigma$ uncertainty estimated using the jackknife. We found weak redshift evolution and hence
  our simulation at fixed redshift should provide a good description of the sample despite its
  non-negligible redshift range.}
\label{fig:wp_zbin}
\end{figure}

We use $N$-body simulations in order to investigate systematic effects and estimate various possible
systematic corrections. We use an $N$-body simulation run using the TreePM method \citep{Bagla2002,
  White2002, Reid14}, provided by Martin White. We are using 10 realizations of this simulation
based on the $\Lambda$CDM model with $\Omega_m= 0.292$ and $h=0.69$. These simulations are in a
periodic box of side length 1380$h^{-1}$Mpc and $2048^3$ particles. A friend-of-friend halo
catalogue was constructed at effective redshift of $z=0.55$. This is appropriate for our
measurements since the galaxy sample used has effective redshift of 0.57. We have found weak redshift evolution of the clustering signal as shown in Figure \ref{fig:wp_zbin}. Therefore our simulations at mean redshift without any redshift evolution should provide a good approximation to the original data.

\begin{figure}
\includegraphics[width=0.45\textwidth]{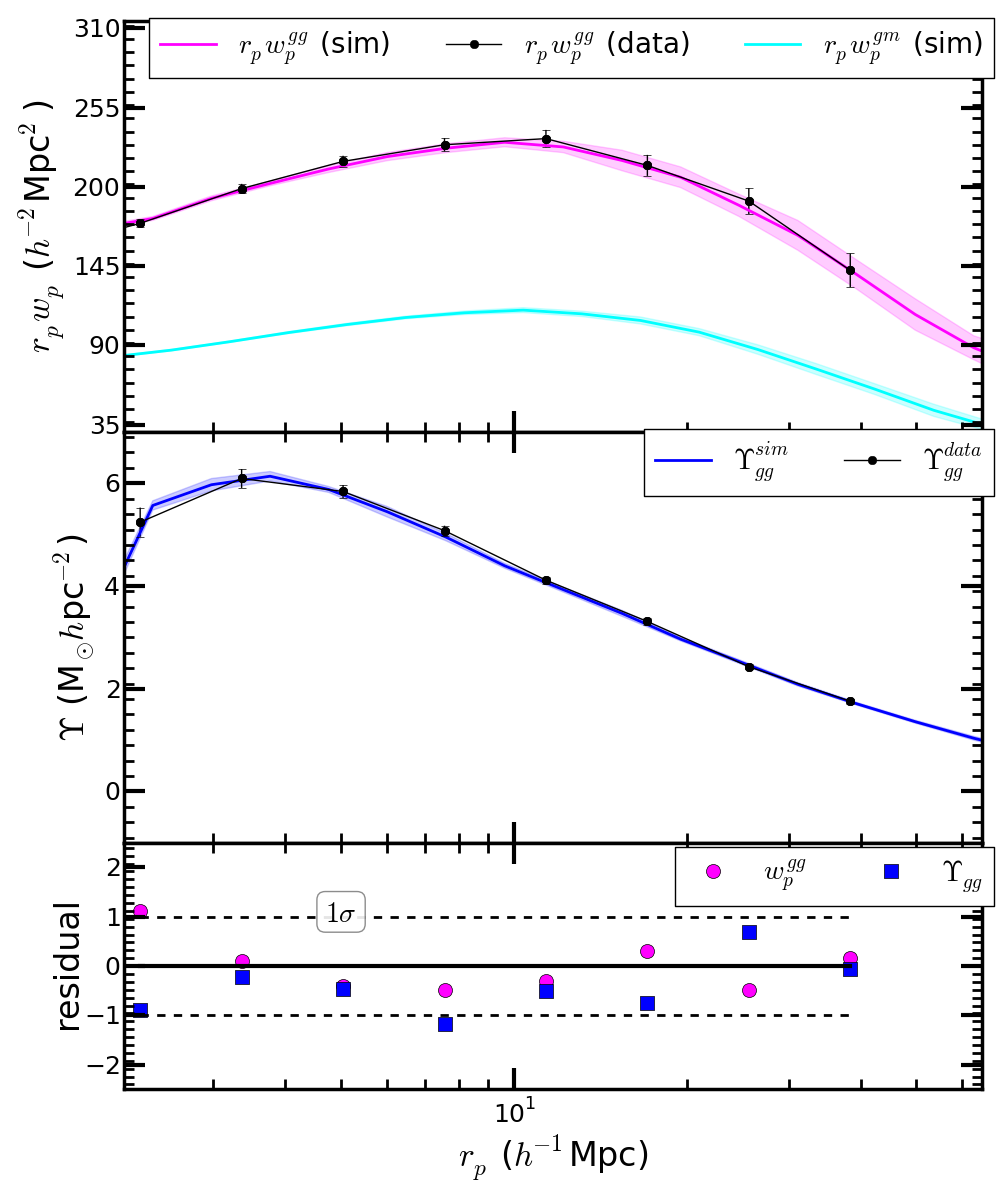}
\caption{The top panel shows the projected correlation function, middle panel shows ADSDs and the bottom panel shows the residual between data and simulation as the function of $r_p$. The black dots are measurement from the BOSS CMASS sample. The red and blue line represents measurements of galaxy-galaxy and galaxy-matter clustering from $N$-body
  simulations. The shaded regions are standard deviation of 10 $N$-body simulations. The red and
  blue points in the bottom panel shows that the residuals (normalized by statistical uncertainties)
  are within $1\sigma$ for both $w_p^{gg}$ and $\Upsilon_{gg}$ respectively. This shows that
  our simulations and observations gives consistent results. }
\label{fig:wp-ups-sim}
\end{figure}

The Halo Occupation Distribution
\citep[HOD;][]{Peacock2000,Seljak2000,Benson2000,White2001,Berlind2002,Cooray2002} is used to relate
the observed clustering of galaxies with halos measured in the $N$-body simulation. The HOD model
used was proposed in \cite{White2011} to populate the halo catalogue with galaxies. The number of
galaxies hosted in each halo is a function of halo mass; halos can host central and satellite
galaxies.  The occupation distributions are
\begin{eqnarray}
\left\langle N_{\rm cen} \right\rangle_M &= \frac{1}{2}  \mathrm{erfc}\left( \frac{\ln (M_{cut}/M)}{\sqrt{2}\sigma}\right) \, , \nonumber \\
\left\langle N_{\rm sat}\right\rangle_M &= \left\langle N_{\rm cen}\right\rangle_M \left(  \frac{M-\kappa M_{cut}}{M_1}\right)^\alpha \,
\label{eqn:HOD}\, ,
\end{eqnarray}
where $\left\langle N_{\rm cen}\right\rangle_M$ is the average number of central galaxies for a
given halo mass $M$ and $\left\langle N_{\rm sat}\right\rangle_M$ is the average number of
satellites galaxies. The HOD parameters we used\footnote{These HOD
  parameters were obtained by carrying out a fit to the projected clustering signal, $w_{\rm p}$
  measured in Section~\ref{sec:ups_gg}, and the CMASS galaxy abundance using the analytical halo
  model framework \citep{vdBosch:2013, More:2013, Cacciato:2013}. The analytical halo model
  developed in these papers accounts for halo exclusion, radial dependence of halo bias, and the
  residual redshift space distortions due to finite extent of the line-of-sight integration used to
  compute $w_{\rm p}$. We refer the reader to these papers and to \citet{More:2015} for the details
  of the modelling procedure. Since the quoted HOD corresponds to halos with an
  overdensity of 200 with respect to the background, while the halos in
  simulations were obtained using the FOF algorithm, we used the relation
  between FOF halo masses and SO halo masses derived in \citet{More:2013} for
  idealized Navarro-Frenk-White halos. While this correction is not perfect in
  practice \citep{More:2015}, it is sufficient to describe the
  clustering signal observed in the data.} are ($M_{cut}=1.77\times 10^{13} M_\odot/h,
M_1=1.51\times 10^{14} M_\odot/h, \sigma=0.897, \kappa=0.137, \alpha=1.151$).
We have populated central galaxies at the center of the halos. The satellite galaxies are populated
with radius (distance from the central galaxy) distributed as
per the NFW profile out to $r_{200}$, and the direction is chosen randomly with a uniform
distribution assuming satellites are spherically distributed. It is a good approximation because only $\sim 10$\% of the galaxies are satellites.
The central galaxies are each assigned the same velocity as their halo. The satellite galaxies are assigned
velocities which are normally distributed, with mean as the halo velocity and dispersion the same as the halo velocity dispersion. 

We find that the clustering measurement between our $N$-body simulation and measurement agrees within $1\sigma$. 
The top panel of Figure~\ref{fig:wp-ups-sim} shows the projected galaxy-galaxy (red) and
galaxy-matter (blue) correlation functions. The shaded regions are the standard deviation of 10 $N$-body mocks. The galaxy-galaxy projected correlation function measured from data shown in black points agrees quite well with the one measured from $N$-body simulation. 
The projected galaxy-galaxy correlation is used to measure $\Upsilon_{gg}$, which is shown in the
middle panel of Figure~\ref{fig:wp-ups-sim}. The projected galaxy-matter cross-correlation function
is used to compute $\Upsilon_{gm}$, which is shown in Figure~\ref{fig:ups-gm-data}. The measurement
of $\Upsilon_{gm}$ from the $N$-body simulation (using the $w_p$-based HOD parameters) and data agrees very well. We used this HOD populated galaxy sample to compute some of the systematic corrections on our measurement (see Section \ref{sec:systematic} for details). Such correction will have an error due to the uncertainty in HOD parameters. Since the HOD parameters are very well constrained due to small error on clustering measurements we expect such error to be small compared to the correction.


\section{Potential Systematics} 
\label{sec:systematic}

We investigate various possible systematic effects which can affect our measurement of $E_G$. We
will show that some of them are negligible and have computed corrections for others. These
corrections are small compared to the statistical uncertainty on the measurement. We have
applied these systematic corrections to our final measurement. An alternative approach to account
for these systematic shifts is to apply them to the theoretical prediction as shown in \cite{Leonard2015}.

\subsection{Scale dependence of bias}
\label{sec:cb}
Galaxies are formed in dark matter halos, which makes the clustering amplitude of galaxies biased
compared to that of dark matter. The massive galaxies used in our analysis are a highly biased
sample. It has been shown that a linear bias model fails to match the observations and simulations at small scales \citep{Saito2014}. $E_G$ is constructed in such a way
that it is independent of linear bias. However, the redshift space distortion parameter $\beta$ is
computed using the linear bias whereas the projected correlation function is calculated at smaller
scales, where the bias is scale-dependent, the causing  bias factor in $E_G$ to not completely
cancel. To correct for this factor, we compute the correction factor
$C_b(r_p)=\Upsilon_{gg}(r_p)/b\Upsilon_{gm}(r_p)$ the from mock catalogs in Section~\ref{sec:nbody}, as proposed by \cite{Reyes2010}.
The correction factor $C_b(r_p)$ is shown in Figure~\ref{fig:sys-corr}, where linear bias $b=1.95$
was measured in RSD analysis of our N-body simulation using RSD model of \cite{Alam2015}. The
$\Upsilon_{gg}$ and $\Upsilon_{gm}$ for this correction is computed using 10 N-body simulations
shown in Figure~\ref{fig:wp-ups-sim}. The scale-dependent bias correction $C_b$ has maximum value of 8\% at 8 h$^{-1}$Mpc. It is important to note that our scale-dependent bias correction does not
approach 1 at the large scales used for this analysis, contrary to our expectations. This is because our largest scale is still in quasi-linear regime and hence the scale-dependent bias correction does not approach 1. 

\subsection{Difference in Lensing and Clustering Window}
\label{sec:cwin}
$E_G$ includes the ratio of the galaxy-galaxy auto-correlation with galaxy-matter cross-correlation
measured from lensing. The galaxy-galaxy auto-correlation is measured in redshift space with a
top-hat window. The  galaxy-matter cross-correlation, on the other hand, is measured with very broad
lensing window that washes out the effect of redshift space distortions and behaves differently with
line of sight separation compared to a top-hat. To correct both of these effects, we use the window
function correction ($C_{win}$). The window function correction consists of two parts $C_{win}=
C_{RSD} C_{integration}$, where the $C_{RSD}$ is the ratio of $\Upsilon_{gg}$ in redshift and in
real space. We compute this correction by evaluating the clustering in real and in redshift space
from 10 N-body simulations. 

The motivation for the $C_{integration}$ correction is as follows. When making our theoretical
predictions for $E_G$, we begin from a 3D model that we assume is projected to 2D in the same way
for both galaxy-galaxy and galaxy-matter correlations, using $ \int_{-\Pi_{\rm max}}^{\Pi_{\rm max}} \xi_{gg,gm}(r_p,\pi) \,d \Pi$. This is a top-hat window with a hard cutoff at $\Pi_{\rm max}$. However, reality provides us with a lensing shear signal that is projected using a non-top-hat window. It is not truncated at $\Pi_{\rm max}$, and it is not flat like a top-hat. We must consider a more generalized situation with some window function $W(\Pi)$ using $ \int_{-\Pi_{\rm max}}^{\Pi_{\rm max}} \xi_{gm}(r_p,\Pi) W(\Pi) \,d \Pi$. Thus, the ratio of quantities used to construct $E_G$ should differ slightly from the theory prediction that assumes a top-hat window for both. The correction factor by which we should multiply our observed $E_G$ (before comparing with
theory) is $C_{integration}$, the ratio of $\Upsilon_{gm}^{(top)}$ with a top-hat window and
$\Upsilon_{gm}^{(win)}$ with the lensing window as determined by the source and lens redshift distribution ($C_{integration}= \Upsilon_{gm}^{(top)}(r_p)/\Upsilon_{gm}^{(win)}(r_p)$).
The ``top'' version is in the numerator because we construct $E_G$ with the real lensing data, i.e., we implicitly computed and used the ``win'' version in the data. We want to divide that out and replace it with the ``top'' version when comparing with the theory. The lensing window can be written as follows:
\begin{align}
W(\Pi = & \chi(z_m)-\chi(z_l))  = \frac{1}{N}\int  dz_l P_{lens}(z_l) \\ 
     & \int_{z_m}^{\infty} dz_s P_{src}(z_s) \Sigma_{cr}^{-2}(z_l,z_s) 
     \frac{\Sigma_{cr}(z_l,z_s)}{\Sigma_{cr}(z_m,z_s) }  \nonumber  \\
     N =& \int  dz_l P_{lens}(z_l) \int_{z_m}^{\infty} dz_s P_{src}(z_s) \Sigma_{cr}^{-2}(z_l,z_s)    
\label{eq:lens-win}
\end{align}

\noindent where the $\Sigma_{cr}(z_1,z_2)$ is given in Eq.~\eqref{eq:sig_cr}. Currently our theory
assumes that we simply take the galaxy matter cross-correlation $\xi_{gm}$, which is the correlation
function between matter at that lens redshift and the lens galaxy position, and projected along the
line of sight with a top-hat window.  But in practice, if you have matter that is correlated with
the lens but not exactly at the lens redshift, then the shear for that source is determined by
$\Sigma_{cr}(z_m, z_s)$, which varies along the line of sight as $z_m$ is closer to or farther away
from the lens. The  $(z_l,z_m,z_s)$ are the redshifts of lens, matter and source respectively. The
innermost integral is to account for the fact that matter at $z_m$ will lens all the sources behind
matter ($z_s>z_m$) and $P_{src}(z_s)$ is the redshift distribution of source in our sample. The
outer integral is to account for the fact that we have a distribution of lens given by
$P_{lens}(z_l)$ which should be integrated over. The lens redshift sets the zero point of the
line-of-sight separation ($\Pi$) for the galaxy-matter cross-correlation.  $\Pi$ is the comoving
distance between matter and lens ($\Pi = \chi(z_m)-\chi(z_l)$). Here $\chi(z)$ is the comoving distance to the redshift $z$. We
compute $W(\Pi)$  with the CFHTLenS source redshift distribution and CMASS lens redshift
distribution. We use the N-body simulation to estimate the galaxy-matter cross-correlation and apply
$W(\Pi)$ in order to compute the projected correlation function with the lensing
window. Figure~\ref{fig:sys-corr} shows the $C_{win}$ correction we have computed for our sample. It
is below 8\% at the scales of interest.

\subsection{Different redshift weighting of lensing and clustering}
\label{sec:cz}
\begin{figure}
\includegraphics[width=0.5\textwidth]{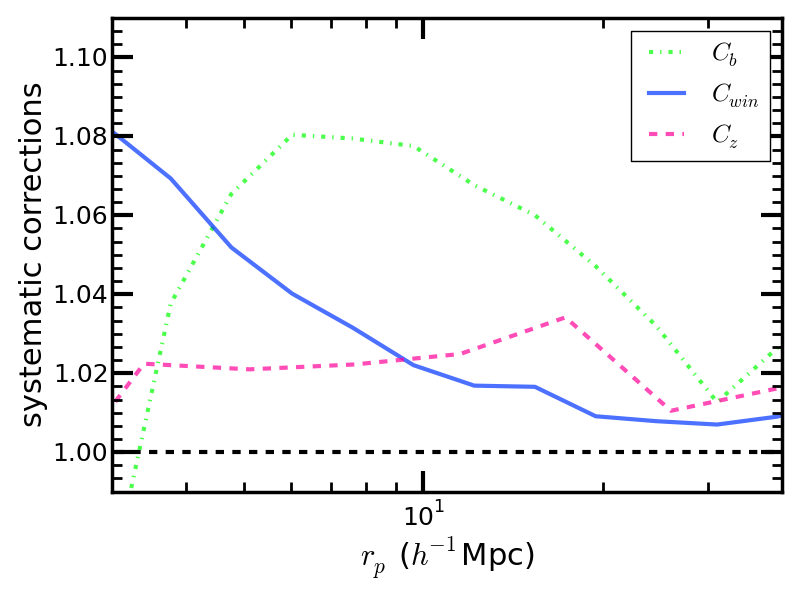}
\caption{Figure shows the systematic corrections estimated for $E_G$ . The green dashed-dotted line is the
  correction for the scale dependent bias ($C_b$; see section \ref{sec:cb}). The blue solid line is for the correction due to
  difference in the clustering and lensing radial windows ($C_{win}$; see section \ref{sec:cwin}). The magenta dashed line is for the
  correction due to difference in redshift weighting of clustering and lensing ($C_z$; see section \ref{sec:cz}). 
}
\label{fig:sys-corr}
\end{figure}

The weighting of the lensing signal averaged over redshift depends on the number of source
galaxies behind a lens galaxy as a function of redshift. Within an annulus of fixed transverse separation, the galaxies at lower redshift have a higher number of sources behind them compared to galaxies at higher redshift. Also, clustering and lensing signal have different redshift weights. Specifically, our clustering measurement uses the weight $w_{\rm tot}$ defined by \eqref{eq:cmass_weight} for each galaxy, while the lensing measurement uses $w_{{\rm tot},l}\sum{w_s}\Sigma_{\rm cr}^{ls\ -2}$ for each galaxy, where the summation runs over lens-source pairs for a given lens galaxy. This
makes the effective redshift of the lensing measurement different from the effective redshift of clustering measurement. 
In order to correct for this difference in redshift,
we compute the multiplication factor $C_z= \Upsilon_{gg}^{clust} / \Upsilon_{gg}^{lens}$. Here $\Upsilon_{gg}^{lens}$ is galaxy-galaxy clustering signal obtained with lensing weight as the function of redshift including both the annulus factor and the redshift weight  and $\Upsilon_{gg}^{clust}$ is galaxy-galaxy clustering signal obtained without lensing weight. This shifts the effective redshift of the
clustering signal to the effective redshift of lensing signal. Figure \ref{fig:sys-corr} shows the $C_{z}$ correction we have computed for our sample. It is at the level of 2\%.

\subsection{Effects of Intrinsic Alignments}
When photometric redshift errors cause galaxies that are at the lens redshift to be included in the
source sample for the lensing meausurement, the lensing measurement can be contaminated by intrinsic
alignments \citep[for reviews, see][]{2014arXiv1407.6990T,2015SSRv..193....1J} of the false sources
towards the lenses.  The majority of the large-scale intrinsic alignment signal is carried by red
galaxies \citep[e.g.,][]{2007MNRAS.381.1197H}, for which the linear tidal alignment model \citep[e.g.,][]{2004PhRvD..70f3526H} provides a reasonable
large-scale description that matches observations of this effect \citep[e.g.,][]{2011A&A...527A..26J}, and for which there are various descriptions on small scales
\citep[e.g.,][]{2015JCAP...08..015B}.  In brief, pressure-supported galaxies form in primordial tidal fields due to large-scale
structure; these tidal fields have stretching axes
that point towards over-densities, and as a result the galaxies that form in those tidal fields also become radially aligned towards
overdensities. This manifests as a negative galaxy-galaxy lensing signal, with
intrinsically-aligned galaxies pointing radially towards the overdensities that our lens galaxies
trace, reducing the
measured $\Upsilon_{\text{gm}}$.  However, our removal of small-scale information partially
mitigates the intrinsic alignment effect, which scales with separation roughly in the same
  way as the matter correlation function.  Also, the photometric redshifts in
CFHTLenS are good enough to reduce the contamination by physically-associated galaxies to a very low
level, as demonstrated by the fact that we found a boost factor consistent with 1 to within 1.6 per
cent at $r_p \simgt 1.5~h^{-1}{\rm Mpc}$ (Section~\ref{sec:measurement}).

To estimate the magnitude of possible intrinsic alignment contamination in the galaxy-galaxy lensing
signal in this work, we need several pieces of information.  The first is the average intrinsic
shear for galaxies that are within $\sim 100h^{-1}$Mpc of the CMASS galaxies in our source sample.
To estimate this, we use the average intrinsic shear of LOWZ galaxies as a function of $r_p$, from
\cite{2015MNRAS.450.2195S}.  We then use the redshift evolution of the linear alignment model to decrease this by a
factor of 0.85 to go to the CMASS redshift \citep{2004PhRvD..70f3526H}, and by a factor of ten to account for the fact that the
sources that are used here are on average about eight times fainter than LOWZ galaxies.  The factor
of ten arises because intrinsic
alignments are consistent with a slightly steeper than linear scaling with luminosity for red galaxies, as
  determined empirically in \citealt{2011A&A...527A..26J}
and \citealt{2015MNRAS.450.2195S}, though this is
an extrapolation below the luminosity range in which their measurements exist.  This gives an estimate of
$\langle \gamma_{\rm int}\rangle$ for the source sample used in this work.  

In cases where there
  are many satellite galaxies at the lens redshift included in the ``source'' sample, and thus a
  boost factor substantially in excess of 1, it is common practice to assume that only those excess
  galaxies are intrinsically aligned \citep[e.g.,][]{2012JCAP...05..041B}. In our case, there are
  essentially no ``excess'' galaxies, but we still must assume that non-excess galaxies near the
  lenses are intrinsically aligned.  To account for this, we calculate the
fraction of sources that are within $100h^{-1}$Mpc along the line-of-sight from a typical CMASS
galaxy, given the source $p(z)$. For example, for a fixed redshift $z_{\rm lens}$,
\begin{equation}
f_{\rm local}(z_{\rm lens}) = \int_{z_{\rm lower}}^{z_{\rm upper}} \mathrm{d}z \, p(z)
\end{equation}
where the lower and upper limits of integration are defined by finding the redshift corresponding to
$\pm 100h^{-1}$Mpc separations from the lens redshift, and we assume the source $p(z)$ are
normalized to integrate to unity over all redshifts  
We average the $f_{\rm local}(z_{\rm lens})$ estimates over the lens redshift distribution. This
average fraction $\langle f_{\rm local}\rangle $ is approximately $0.05$.
Finally, we compare $\langle \gamma_{\rm int}\rangle \langle f_{\rm local}\rangle $, which is the total estimated intrinsic
alignments contamination to the shear, with the measured shear.  The estimated contamination has a
maximum value (as a function of $r_p$) of 
$1$ per cent of the measured shear, or at most $0.1\sigma$.  Even if some of the above assumptions
are incorrect by a factor of two, we conclude that we can safely ignore intrinsic alignment
contamination in our measurement, particularly given that (a) the estimates from \cite{2015MNRAS.450.2195S} were for red
galaxies, and many of the sources are blue galaxies; and (b) the redshift-dependent lens-weighting
will suppress the contributions from these more ``local'' galaxies that may be intrinsically aligned.

\subsection{Fingers of God in RSD $\beta$}

Galaxies that are satellites orbitting within the same halo exhibit random motions due to the
gravitational potential well of the halo. This can cause galaxies at the same
line of sight distance to have different redshifts, so that they 
 appear to be spread out into a very elongated structure along the line of sight. This effect is known as fingers of god. We have modeled the fingers of god by introducing a
parameter $\sigma_{FOG}$, a scale-independent additive term in the velocity dispersion of the Gaussian
Streaming Model (GSM) \citep{ReiWhi11}. \cite{Alam2015} shows that this model recovers the expected parameter for the PTHalo mocks \citep{Manera13} and $N$-body mocks (\citet{Tinker2016}, in prep.).
 This means that our fingers of god modeling is accurate enough for the scales used in our redshift
 space distortion analysis. The expected bias in the measurement of $f$ should be below 2.6\%, much
 smaller than the uncertainty in other measurements, as shown in Figure~\ref{fig:error-decomp}. We  do not expect any extra bias in our $E_G$ measurement from $\beta$.

\subsection{Cosmic Variance}
\label{sec:cv}
\begin{figure}
\includegraphics[width=0.5\textwidth]{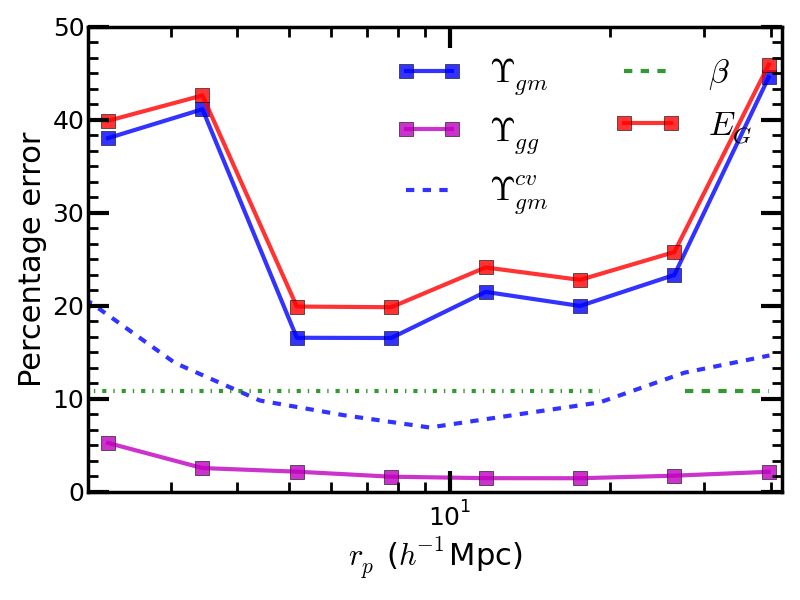}
\caption{Figure shows the percentage error in different components of our measurement. The magenta
  line shows the error in $\Upsilon_{gg}$, which is the lowest. The blue line represents the lensing
  error, and the dashed blue line is an estimate of the cosmic variance due to the finite size of
  the CFHTLenS fields. The green line is the 
  error on the measurement of the RSD parameter $\beta$, and the red line is the combined error on our measurement of
  $E_G$.  Note that $\beta$ is measured only above 26 h$^{-1}$Mpc, shown with a dashed line.}
\label{fig:error-decomp}
\end{figure}

The CFHTLenS is a relatively small area of sky covering 170 square degree. The overlap between
CFHTLenS and BOSS is 105 square degrees. 
This raises the question of our lensing measurement being limited by cosmic variance. We have
estimated the cosmic variance from simulations by dividing our $N$-body simulation into roughly 105
square degree regions at $z=0.57$. The CFHTLenS survey consists of four fields W1--W4 with
overlapping areas of 48, 3, 33, and 21 square degrees with the CMASS DR11 sample respectively. The
areas of each of the four fields are equivalent to areas of squares of side length 176, 44, 146 and
116 h$^{-1}$Mpc at redshift $z=0.57$. We tried to mimic this in our simulations by having four
square with equivalent areas separated by 100 h$^{-1}$Mpc from each other. Also, we limit our
galaxies in the simulations along the $z$ axis within 600 h$^{-1}$Mpc in order to have a
line-of-sight extent equivalent  to CMASS sample between redshift of 0.43 and 0.7. However, matter
particles are used for the full extent of the periodic box along $z$ axis to account for the broad
extent of the lensing kernel. We created 89 such realizations and computed $\Upsilon_{gm}$ for each
realization. The variance of $\Upsilon_{gm}$ from these 89 realizations should give us an estimate
of cosmic variance in our analysis. We have found that the cosmic variance is comparable to the
lensing statistical error due to shape noise. We have also looked at the effect of changing the distance between our four sub-fields and found no significant affect in our estimate of cosmic variance. Figure~\ref{fig:error-decomp} shows the percentage error in different component of our measurement. The dashed blue line in the figure represents cosmic variance on $\Upsilon_{gm}$.  We believe that our cosmic variance estimates are underestimated in the largest bins by not incorporating the actual shapes of the CFHTLenS subfields in our simulation, which can reduce the number of large scale modes available and hence increase the error in the largest bins. We estimate error on our lensing measurement using two methods described in section \ref{sec:upgm} and shown in Figure~\ref{fig:Lensing-error}. We found that the difference between the covariance
estimated from MD-Patchy mocks and using a combination of randoms with cosmic variance is within 3 per cent at $r_p\simlt
30h^{-1}{\rm Mpc}$. We use the error estimates from MD-Patchy in our final results.  

\subsection{Calibration Bias}
Biases in the estimation of the ensemble lensing shear from the shape measurements of galaxies are
one of the major systematics in galaxy lensing measurements. The inaccurate modeling of galaxy shape
and PSF could leave both isotropic and anisotropic residuals in the ensemble shear. These residuals
affect our shear measurements and can be parametrized as multiplicative and additive corrections
\citep{Heymans2006}. The shape catalogue of CFHTLenS used in our analysis is corrected for these
effects as described in \cite{Miller:2013}. We have also shown that our HOD tuned to match galaxy
clustering also matches the lensing measurements without any tuning (see Figs.~\ref{fig:ups-gm-data}
and~\ref{fig:wp-ups-sim}), indicating the absence of any statistically significant calibration
bias. Although this is not a perfect test, as differences in cosmological parameters such as
$\Omega_m$ between simulation and reality could potentially absorb such a bias.  Note that
\cite{LiuJ2016} shows that a multiplicative bias could be detected in the faint subset of the
CFHTLenS sample, but this could also be absorbed in many other unknowns and not necessarily hint
towards a need for multiplicative bias correction.

Another possible systematic uncertainty in the lensing measurements is due to the bias in the photo-$z$ estimates. In \cite{Miyatake:2015}, they confirmed that the shift of the lensing signal is within a few percent for the possible range of the photo-$z$ bias $\delta z=\pm 0.02$ \citep{Erben2013}, which is well below the statistical uncertainty in our lensing measurements.

\subsection{Impact of $R_0$}
\label{sec:R0}
The $\Upsilon$ used in the measurement of $E_G$ attempts to remove the information from scales below $R_0$ and makes the measurement insensitive to our lack of understanding of small scale physics. In principle, there are many unknowns at scales below the virial radius, which are sensitive to baryonic physics, feedback models, stochasticity and other details related to galaxy formation. Also, the simulation used in our study lacks all of the above mentioned details and is obtained by populating $N$-body halo catalog using HOD model. This lack of details leads to failure in the description of the clustering and $E_G$ measurement at the smallest scales. It is important to note that the choice of $R_0$  affects the measurement of $E_G$ above $R_0$ by changing the scale dependence of the bias. We remove the impact of $R_0$ from larger scales by applying a correction factor $C_b$ (see section \ref{sec:cb} for details). Therefore measurements of $E_G$ will be insensitive to $R_0$ as long as it is large enough to remove scales which we do not model in our simulations and corrections. Hence the choice of $R_0$ should be based on the minimum scales that our simulations can model for the observed clustering and lensing signal within error. We have found that our simulations can model the clustering and lensing signal down to scales $1.49h^{-1}$Mpc (see Figure \ref{fig:ups-gm-data} and \ref{fig:wp-ups-sim}). Therefore, we used $R_0=1.49h^{-1}$Mpc in this paper.  Note that a higher value of $R_0$ will also degrade the signal to noise ratio of our final measurement. Hence one should  minimize the value of $R_0$.

We have also investigated the correction factor $C_b$ when we use the extreme case of $R_0=0$. This will imply using $\Delta \Sigma$ in place of $\Upsilon$ for the measurement of $E_G$. In order to compute the correction factor for such a scenario, we need to predict the clustering signal down to small scales beyond the capability of our simulations. Based on our HOD modeling we estimated that only ~10\% of CMASS galaxies are satellites. This means that the two-point correlation function at such small scales will be dominated by central-satellite pairs. 
If we assume that satellite galaxies sample NFW (halo density) profile, we can make a reasonable assumption that small scale clustering is given by the NFW profile. We obtained the maximum correction factor $C_b$ to be 6\% which is slightly smaller than the 8\% correction obtained for $\Upsilon$. So, we note that, if small scale clustering is modeled with NFW profile then the scale dependent correction factor $C_b$ obtained using either $\Delta \Sigma$ or $\Upsilon$ are similar. We use $\Upsilon$ in the rest of the paper because this doesn't require any extra assumption about the nature of small scale clustering.

\section{Results} 
\label{sec:results} 

In this section we provide the details of our measurement of $E_G$ and its covariance.

\subsection{Measurement of $E_G(r_p)$}
\begin{figure}
\includegraphics[width=0.5\textwidth]{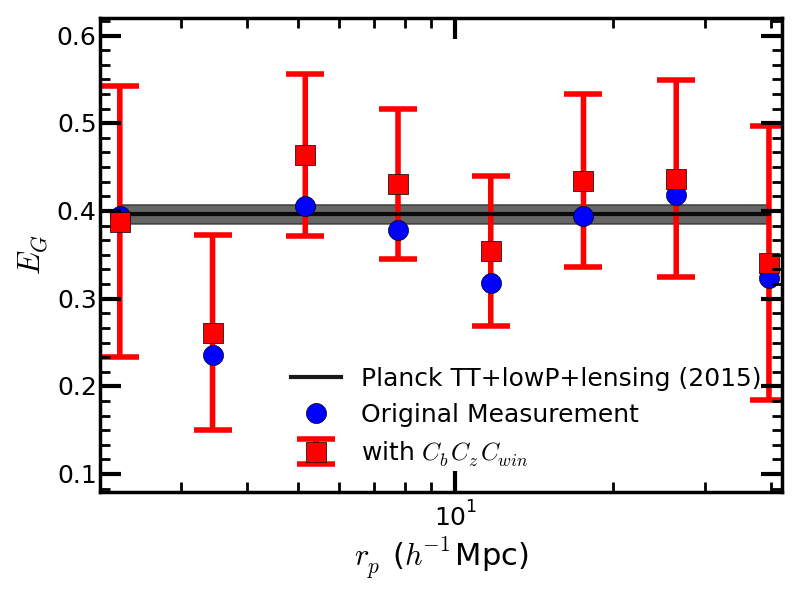}
\caption{The plot shows $E_G$ as a function of $r_p$. The blue points show the raw measurement (without any corrections)  and the red points are the final measurement after multiplying by systematic corrections ($C_b C_z C_{win}$). The black line is the prediction of GR for Planck (2015; TT+lowP+lensing) cosmology with shaded region representing 1$\sigma$ error.}
\label{fig:EG_R}
\end{figure}

$E_G$ is a combination of three different signals that is designed to be more sensitive to the modification of gravity. We have measured
$\Upsilon_{gg}(r_p)$ and $\Upsilon_{gm}(r_p)$ for $2.28<r_p<40h^{-1}$Mpc in 8 logarithmic bins as
described in Section~\ref{sec:measurement}. We combine our measured signal to get $E_G(r_p)$ as in
Eq.~\eqref{eq:EGobservation}, then multiply by $C_b C_{win} C_z$ as in Sec.~\ref{sec:systematic} in
order to correct for differences in how $\Upsilon_{gg}$ and $\Upsilon_{gm}$ are measured that result
in deviation from theoretical predictions. Figure~\ref{fig:EG_R} shows our measurement of
$E_G(r_p)$. The blue (red) points show the measurement before (after) systematic corrections.  The
black line shows the GR prediction and the shaded region is one sigma error according to Planck
(2015; TT+lowP+lensing; \citealt{planck15}).

\subsection{Covariance matrix of $E_G$}
The covariance of $\Upsilon_{gm}$ has been computed as described in
Section~\ref{sec:measurement}. The covariance on $\Upsilon_{gg}$ is obtained using jacknife. The
error on measurement of $\beta$ was obtained as part of the redshift space distortion analysis
described in \cite{Alam2015}. We compute the diagonal error on $E_G$ by adding the errors from $\Upsilon_{gm}$, $\Upsilon_{gg}$ and $\beta$ in quadrature as
\begin{equation}
{\sigma_E}_G(r_p) = E_G(r_p) \sqrt{\left(\frac{{\sigma_\Upsilon}_{gm}}{\Upsilon_{gm}}\right)^2 + \left(\frac{{\sigma_\Upsilon}_{gg}}{\Upsilon_{gg}}\right)^2 + \left(\frac{\sigma_\beta}{\beta}\right)^2 } \,
\label{eq:sigEG}
\end{equation} 

\begin{figure}
\includegraphics[width=0.5\textwidth]{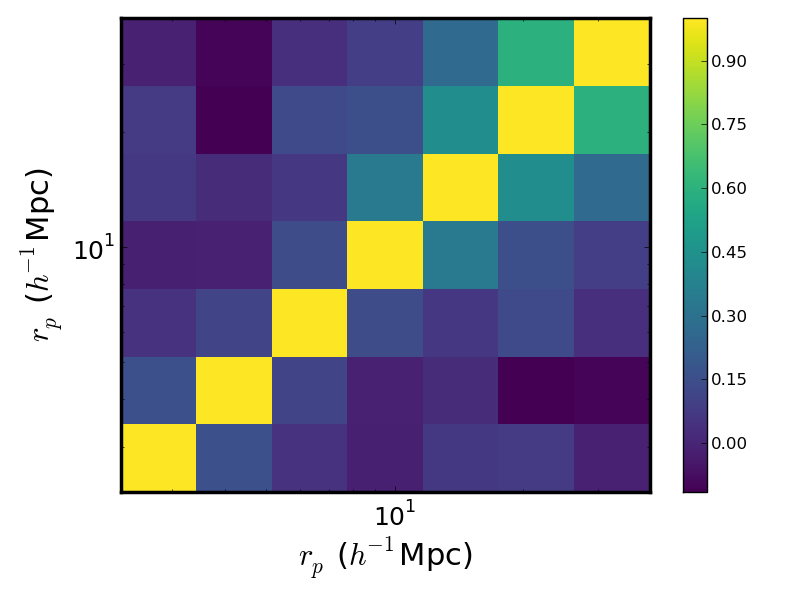}
\caption{{\bf Correlation matrix ($\Psi$) of $E_G$:} We have estimated the covariance of $\Upsilon_{gm}$ using 150 MD-Patchy mocks. This covariance is then used to compute the correlation matrix (see Eq.~\ref{eq:psi} for details). 
 }
\label{fig:EG_cov}
\end{figure}

The above equation also assumes that the signal-to-noise ratio of all quantities is high enough that the error
  distribution on the ratio is a Gaussian. This is a good assumption because the signal in each bin
  is detected at the 5$\sigma$ level or better. We then compute the correlation matrix of $\Upsilon_{gm}$ given by
\begin{equation}
\Psi_{gm}(r_i,r_j)=
\Sigma_{gm}(r_i,r_j)/\sqrt{(\Sigma_{gm}(r_i,r_i)\Sigma_{gm}(r_j,r_j)}
\label{eq:psi}
\end{equation}
where $\Sigma_{gm}(r_i,r_j)$ represents
the covariance matrix of $\Upsilon_{gm}$.  The covariance matrix for $E_G$ is obtained by multiplying the ${\sigma_E}_G$ with the correlation matrix,
\begin{equation}
{\Sigma_E}_G(r_i,r_j)={\sigma_E}_G (r_i) {\sigma_E}_G (r_j) \Psi(r_i,r_j) 
\label{eq:covEG}
\end{equation}
Figure~\ref{fig:EG_cov} shows the correlation matrix of $E_G$ we have measured. We have assumed that
the different components of $E_G$ are independent while estimating the  covariance matrix. It
is a reasonable assumption because the clustering signal $\Upsilon_{gg}$ and lensing signal
$\Upsilon_{gm}$ are integrated along the line of sight and hence will not be correlated with the
redshift space distortions parameter $\beta$. Also, the lensing measurement is dominated by shape noise
with errors at the 20\% level on all scales, whereas the statistical errors on the clustering
measurements are below 5\% at all scales. Moreover the lensing is measured in a very small fraction
of the area used for the clustering measurement.  For both reasons, the clustering and lensing are
independent in our analysis, justifying the use of the $\Upsilon_{gm}$ correlation matrix to estimate the covariance matrix for $E_G$. 

\subsection{Constraint on $E_G$}

\begin{figure}
\includegraphics[width=0.5\textwidth]{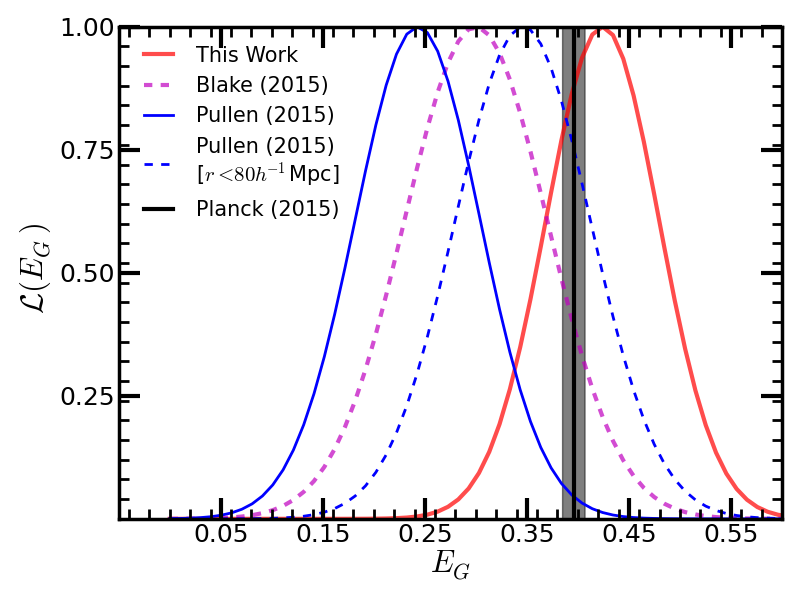}
\caption{Figure shows the one-dimensional likelihood of $E_G$. The red solid line shows our measurement
  $E_G=0.42\pm0.056$. The magenta dashed line is the measurement reported in \citet{Blake2015},
  $E_G=0.30\pm0.07$. The blue solid and dashed lines are measurement from \citet{Pullen2015data} using all scales ($E_G=0.243 \pm 0.061$) and scales below $80 h^{-1}$Mpc ($E_G=0.346 \pm 0.066$) respectively. The black shaded region is the GR prediction for Planck (2015;
  TT+lowP+lensing). Note that \citet{Blake2015} assumes a cosmology with $\Omega_m=0.27$ whereas we
  use the Planck 2015 cosmology. 
}
\label{fig:EG_like}
\end{figure}

We have shown our measurement of $E_G$ in logarithmic bins of $r_p$ in Figure~\ref{fig:EG_R}. We can
obtain a measurement of $E_G$ at an effective average scale by combining the information from all
scales. The constant $E_G$ model can be used in order to obtain the constraint on $E_G$ using our
measurement. We have used measurements between $r_p$ of $5.17h^{-1}$ Mpc and $26.4h^{-1}$ Mpc. The
lower limit is to avoid small scales where systematic corrections become large and baryonic physics
might start to become important~\citep{Mohammed2014}. The upper limit is
determined by the scale at which the systematic correction for the lensing becomes substantial
compared to the lensing signal itself. We fit our measurements of $E_G(r_p)$ with the
full covariance matrix using the model of constant $E_G$, giving $E_G(z=0.57)=0.42 \pm
0.056$. Figure~\ref{fig:EG_like} shows our likelihood for $E_G$ as a red solid line. The black
line and the shaded region are the Planck (2015;TT+lowP+lensing) prediction. The magenta dashed line
is the measurement of \citet{Blake2015}. Our measurement is consistent with the  Planck prediction
and agrees with \cite{Blake2015}. Our final measurement of $E_G$ has an $13\%$ statistical
uncertainty, which is $30\%$ improvement on the previous best measurement at the same redshift
\citep{Blake2015}.  Note that the \citet{Blake2015} measurement is 1.4$\sigma$ from the Planck 2015
cosmology but less than 1$\sigma$ from the WMAP7 cosmology, which is closer to their assumed cosmology.

\section{Discussion}
\label{sec:discussion}

We have analyzed data from CFHTLenS \citep{Heymans2012} and the SDSS-III BOSS DR11 CMASS sample
\citep{Alam2014}. We have measured tangential shear by cross-correlating the CFHTLenS galaxy shapes with the
lens sample (CMASS). This produces a measure of the excess surface mass density $\Delta \Sigma$. We
have also measured redshift space galaxy-galaxy clustering ($w_p(r_p)$) and the redshift space distortions parameters ($\beta$). All of these measurements are tested for various systematics as
described in Section~\ref{sec:measurement}. We then cast these measurements in terms of the annular
differential surface densities \citep[ADSDs;][]{Baldauf2010} to suppress the small scale information.  The $\Upsilon_{gg}$ is defined to match the kernel with $\Upsilon_{gm}$. These
measurements are then combined to estimate $E_G(r_p)$ (see Figure~\ref{fig:EG_R}). We have also
estimated the covariance on our measurements by combining the covariance of $\Upsilon_{gm}$ with the
diagonal error on $\Upsilon_{gg}$ and $\beta$ in quadrature (see Figure~\ref{fig:EG_cov}). The
scale-averaged measurement of $E_G$ is obtained by fitting a constant $E_G$ model. We have also
considered potential systematic errors that can affect our measurements of $E_G$ and computed
possible corrections or provided  upper limits (see Section~\ref{sec:systematic}). We finally report
$E_G(z=0.57) = 0.42 \pm 0.056$ (13\% error) compared to the $\Lambda$CDM prediction of $E_G=0.40$
using the \cite{PlanckI} cosmology.

Our measurements are completely consistent with the prediction of $\Lambda$CDM, and provide a
non-trivial test of GR at cosmological scales by virtue of probing both metric potentials. The first
measurement of $E_G$, reported in \citet{Reyes2010} at redshift of 0.32, was also consistent with
$\Lambda$CDM. A more recent measurement was reported in \citet{Blake2015} at redshifts  0.32 and 0.57. We improve on the measurement of \citet{Blake2015} by about $30\%$ in precision at redshift of 0.57. 
This improvement largely comes from the fact that we are using the BOSS DR11 sample, which has more
data compared to the BOSS DR10 sample used by \citet{Blake2015}, and from the improved precision on
$\beta$ measurement, which we obtained using a different perturbation theory template.  A similar measurement was first proposed in \citet{Pullen2015} and measured in
\citet{Pullen2015data} by replacing the gravitational lensing shear estimated using galaxies with
CMB lensing. This is a complimentary measurement to ours by virtue of probing different scales with
different systematics. \citet{Pullen2015data} reported $E_G=0.243 \pm 0.061$ using scales upto $150h^{-1}$Mpc and $E_G=0.346 \pm 0.066$ using scales upto $80h^{-1}$Mpc. Our measurement is consistent with the measurement of $E_G$ using CMB lensing at  small scales ($r<80h^{-1}Mpc$). But it shows 2.8$\sigma$ tension when compared with their final results, which include large scales. This might indicate that these measurements have reached a limit where observational systematics are approaching the statistical uncertainty, and
  future surveys will require improved analysis methods.

We are entering the golden age of precision cosmology with much bigger and deeper surveys. For
example, we have HSC, KIDS and DES taking data now, and LSST, WFIRST and Euclid happening in the
next decade. The next generation surveys will provide an unprecedented  handle on statistical
errors, which necessitates a much better understanding of systematic errors. Using future surveys,
we will be able to measure $E_G$ much more precisely at multiple redshifts and over a wide range of
scales. Such measurements will enable us to test the predictions of the $\Lambda$CDM model of
structure formation as a function of scale and time, which might provide key insights into dark
energy, dark matter, and the theory of gravity.

\section*{Acknowledgments}

We would like to thank Sukhdeep Singh for many insightful discussion during the course of this project. We would also like to thank anonymous referee for useful comments.
SA and SH are supported by NASA grants 12-EUCLID11-0004 and NSF AST1412966 for this work.  SH is also supported by DOE and NSF AST1517593. 
HM acknowledges the support of Japan Society for the Promotion of Science (JSPS) Research Fellowships for Young Scientists and the Jet Propulsion Laboratory, California Institute of Technology, under a contract with the National Aeronautics and Space Administration.
RM acknowledges the support of the Department of Energy Early Career Award program.

SDSS-III is managed by the Astrophysical Research Consortium for the Participating Institutions of the SDSS-III Collaboration including the University of Arizona, the Brazilian Participation Group, Brookhaven National Laboratory, Carnegie Mellon University, University of Florida, the French Participation Group, the German Participation Group, Harvard University, the Instituto de Astrofisica de Canarias, the Michigan State/Notre Dame/JINA Participation Group, Johns Hopkins University, Lawrence Berkeley National Laboratory, Max Planck Institute for Astrophysics, Max Planck Institute for Extraterrestrial Physics, New Mexico State University, New York University, Ohio State University, Pennsylvania State University, University of Portsmouth, Princeton University, the Spanish Participation Group, University of Tokyo, University of Utah, Vanderbilt University, University of Virginia, University of Washington, and Yale University.

This work is based on observations obtained with MegaPrime/MegaCam, a joint project of CFHT and CEA/IRFU, at the Canada-France-Hawaii Telescope (CFHT) which is operated by the National Research Council (NRC) of Canada, the Institut National des Sciences de l'Univers of the Centre National de la Recherche Scientifique (CNRS) of France, and the University of Hawaii. This research used the facilities of the Canadian Astronomy Data Centre operated by the National Research Council of Canada with the support of the Canadian Space Agency. CFHTLenS data processing was made possible thanks to significant computing support from the NSERC Research Tools and Instruments grant program.


\bibliography{Master_Shadab,papers}
\bibliographystyle{mnras}

\label{lastpage}

\end{document}